# POEM: Pricing Longer for Edge Computing in the Device Cloud
# A Brief Technical Report

Qiankun Yu, Jigang Wu, and Long Chen

School of Computer Science and Technology,

Guangdong University of Technology, Guangzhou, China

yuqiankun1992@foxmail.com, asjgwucn@outlook.com, lonchen@mail.ustc.edu.cn

**Abstract.** Multiple access mobile edge computing has been proposed as a promising technology to bring computation services close to end users, by making good use of edge cloud servers. In mobile device clouds (MDC), idle end devices may act as edge servers to offer computation services for busy end devices. Most existing auction based incentive mechanisms in MDC focus on only one round auction without considering the time correlation. Moreover, although existing single round auctions can also be used for multiple times, users should trade with higher bids to get more resources in the cascading rounds of auctions, then their bud- gets will run out too early to participate in the next auction, leading to auction failures and the whole benefit may suffer. In this paper, we formulate the computation offloading problem as a social welfare optimization problem with given budgets of mobile devices, and consider pricing longer of mobile devices. This problem is a multiple-choice multi-dimensional 0-1 knapsack problem, which is a NP-hard problem. We propose an auction framework named MAFL for long-term benefits that runs a single round resource auction in each round. Extensive simulation results show that the proposed auction mechanism outperforms the single round by about 55.6% on the revenue on average and MAFL outperforms existing double auction by about 68.6% in terms of the revenue.

## 1 Introduction

In the past few years, despite the increasing capabilities of mobile devices including smart phones, Internet of Things (IoT) devices, and wearable devices, resource requirements for mobile applications can often transcend the computation of a single device [1{4]. Therefore, mobile cloud computing is proposed to offload tasks to remote cloud for execution [5{10], though it may introduce longer delay and user experience may suffer. Moreover, long distance tele-

This work has been accepted by the 18th International Conference on Algorithms and Architectures for Parallel Processing, it is not the final version. This work is used for quick sharing of academic findings. The copyright is held by the corresponding copyright holders. Please cite:
Qiankun Yu, Jigang Wu, and Long Chen, “POEM: Pricing Longer for Edge Computing in the Device Cloud”, the 18th International Conference on Algorithms and Architectures for Parallel Processing (ICA3PP), Guangzhou, China. Doi:10.1007/978-3-030-05057-3_28, ICA3PP 2018, LNCS 11336.

communication will consume more energy. In recent work, multiple access mobile edge computing has been proposed as a promising technology to bring computation services close to end users, by making good use of edge cloud servers. There are three types of architecture used in edge computing [11]: edge server, coordinator device, and device cloud. This paper uses the third architecture, as shown in Fig. 1. The computation offloading [12] can be performed in Mobile Device Clouds (MDC) [13-16], which use idle resources of nearby mobile devices to execute tasks. However, mobile devices that provide idle resources may also incur extra cost to themselves, which should be monetary compensated.

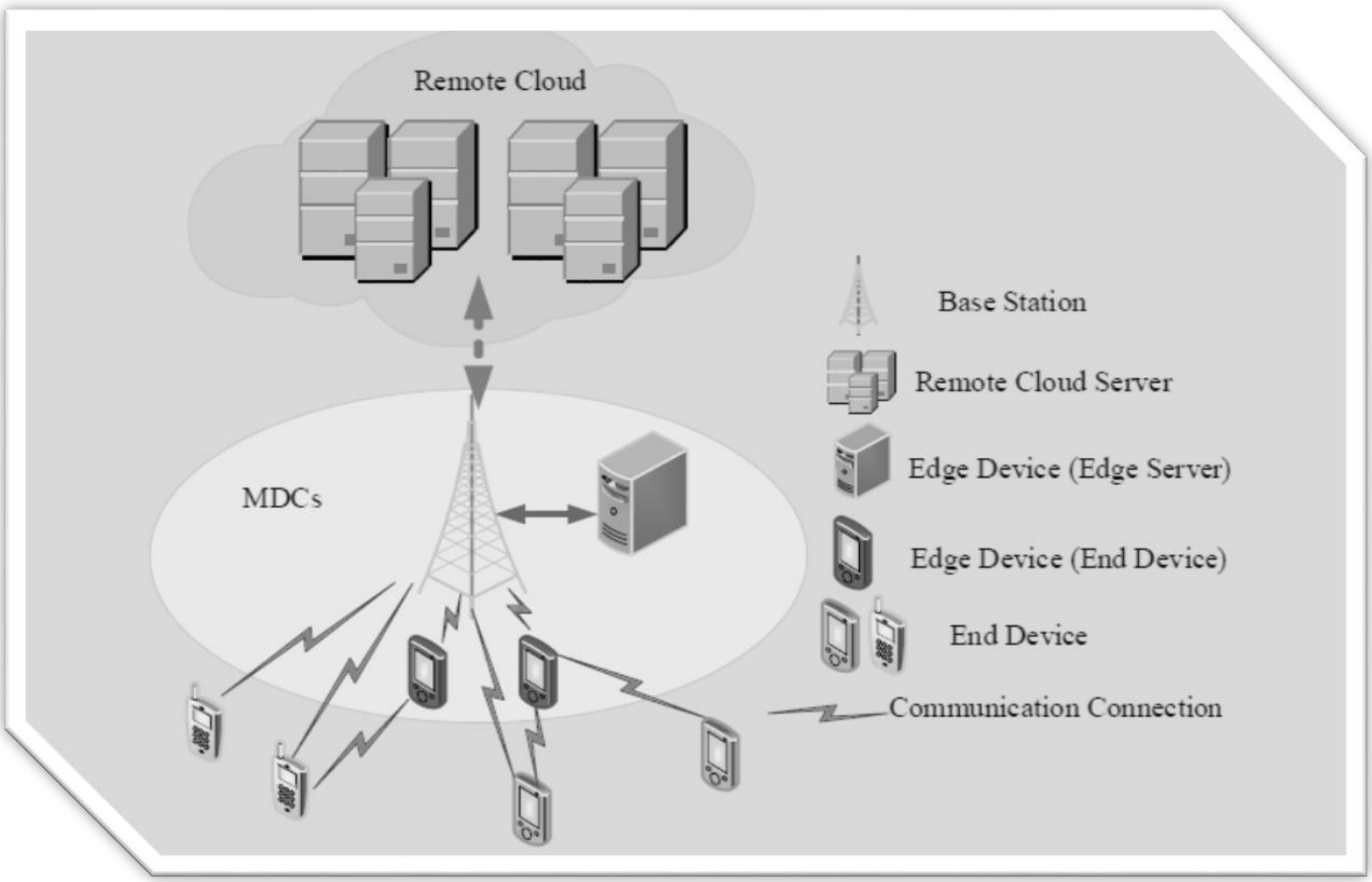

Fig. 1 Network Architecture

To encourage more devices sharing their idle resources, several prior works have been done in MDC. Miluzzo et al. [17] proposed an incentive scheme in MDC. However, this scheme ignored the resource requirements of tasks. Song et al. [18] designed a non-competitive pricing mechanism, with a bill backlog threshold. If a device exceeds the threshold, it can reduce its bill backlog by providing services for others. Otherwise it will not be able to get the service. However, they do not consider whether the device has sufficient resources to provide services for others or not. Wang et al. [19] proposed a Stackelberg Game approach for cooperative application execution in mobile cloud computing. However, they do not consider that the mobile device is heterogeneous, different mobile devices may have different processing power levels and energy consumption levels. Therefore, the payment of the tasks should be different. In recent studies, auction has been widely used as one of the most popular incentive schemes in many areas, such as virtual machine allocation [20, 21] and wireless spectrum allocation [22, 23]. The celebrated VCG mechanism [24] is a well-known type of auction. It is essentially the only type of auction that simultaneously guarantees both truthfulness and absolute economic efficiency. Li et al. [25] proposed an online spectrum auction framework. This mechanism can also be used in MDC's resource allocation, but buyer's budget constraints are not considered. Jin et al. [26] designed an incentive compatible auction mechanism for cloudlet resource sharing in mobile cloud computing. However, this mechanism uses a one-to-one match and assumes that the resource requirements are homogeneous. In this work, we consider a seller can serve multiple buyers and the resource requirements of buyers are heterogeneous. Wang et al. [27] designed an efficient auction mechanism to solve the task assignment problem in MDC. However, this auction mechanism assumes that every buyer must be allocated to resources. In this work, we consider that resources are limited and cannot ensure that every buyer can be allocated resources. In MDC, existing auction mechanisms only focus on a single-round auction [26, 27]. In many cases, we need multiple rounds of auctions. For example, it takes half an hour to play a game, and utilize the unused resources of the nearby mobile devices. The demands for resources are different as the game goes on. It is impossible for a resource auction to be completed just once. Although existing single round auctions can also be used for multiple times; the user's budget constraints should be considered. The budget is the total amount of money a buyer could pay; it plays a key role in designing periodical auctions.

Table 1. Example of Multiple Single-round Auctions

| | budget | l = 1 | l = 2 | l = 3 | l = 4 | l = 5 | l = 6 |
|---|---|---|---|---|---|---|---|
| u1 | 15 | 3 | 4 | 3 | 2 | 1 | 1 |
| u2 | 9 | 4 | 5 | 0 | 0 | 0 | 0 |
| u3 | 10 | 5 | 5 | 0 | 0 | 0 | 0 |
| utility | - | 9 | 10 | 3 | 2 | 1 | 1 |

Table 2. Example of One Improved Auction Framework

| | budget | l = 1 | l = 2 | l = 3 | l = 4 | l = 5 | l = 6 |
|---|---|---|---|---|---|---|---|
| u1 | 15 | 3 | 5 | 4 | 3 | 2 | 1 |
| u2 | 9 | 4 | 2 | 1 | 2 | 1 | 1 |
| u3 | 10 | 5 | 2 | 3 | 2 | 2 | 0 |
| utility | - | 9 | 7 | 7 | 5 | 4 | 2 |

For example, there are three users u1, u2 and u3 competing for two resources. The bidder with the highest bid wins. Their budgets are 15, 9 and 10, respectively. Table 1 is a single round auction cycle multiple times. Each of the columns represents a round of auctions. The winning user is marked with underlines in each round. Bids for users u1, u2 and u3 are 3, 4 and 5 in the first round (l=1), respectively. The winners are users u2 and u3. The utility is 4+5=9 in the first round. The utility here refers to the sum of the winner's bids. The next auction is the same. From Table 1, we can see that user u2 and u3 have high bids to get resources, causing the budget to run out in the previous rounds. As a result, the user u1 has no competition and a lower bid, and the total utility will also be reduced. The total utility here refers to the sum of utility in all round auction, that is 9+10+3+2+1+1=26. Table 2 is an example of one improved auction framework. We punish the winning user to reduce the bid according to his remaining budget in the next round of the auction. In Table 1 and Table 2, the total utility is 26 and 34, respectively. Therefore, the multiple single-round auction can be improved by about 30.7% on the total utility.

Despite long-term and budget constraints are considered in crowdsourcing [28], tasks are homogeneous and a task can be allocated to multiple workers. However, authors in [28] as assumed unlimited resources at workers. In a resource limited MDC, idle mobile devices are unlikely to meet the needs of all users at the same time. So the scheme can't be used directly in MDC, which motivates us to design a long-term auction of multiple rounds with budget constraint. To design effective schemes, the following challenges should be property handled:(1) How to prevent the user's budget from running out prematurely by multi-round auction? (2) How to efficiently allocate resources for different bids of different devices? (3) How to attract more sellers to participate in MDC?

To solve the above challenges, in this paper, we consider Pricing lOnger for Edge coMputing in the device cloud (POEM). We aim to design a long-term auction of multiple rounds. The main features are as follows: (1) the mobile tasks are indivisible and the number of resources (CPU, memory, battery etc.) requirements for a task are different. (2) The number of resources requested by each user is not a fixed value in each round, and the amount of resources provided by the nearby mobile device is also not a fixed value in each round. (3) We punish the winning user to reduce the bid according to its remaining budget in the next round of the auction. In MDC, users could pay high bid to get more resources, then their budget will run out too early to participate in the next auction, leading to auction failure and the whole benefit will suffer.

- Considering the time correlation of resource allocation, we formulate the task offloading problem as an integer linear programming. And we design an MDC Auction Framework for Long-term (MAFL). The next round of genuine bids will be adjusted according to the results of the previous round.
- We consider the user's budget constraints in the MDC. Because it is impossible for users to apply resources without restrictions in the whole auction period.
- Sellers also need to complete their own tasks, and they cannot share their resources without any restriction. Therefore, we consider the constraints of the number of resources shared by sellers in the whole auction period.
- We design a Single Round Mobile Resources Auction (SRMRA) algorithm for comparing purposes with the MAFL. And we demonstrate the performance of the algorithm by proofs and extensive experiments.
- We conduct extensive simulation experiments to demonstrate the performance of our mechanism. MAFL is better than the single round auction SRMRA. MAFL outperforms SRMRA by about 12.2% on revenue when the number of users is 40. MAFL outperforms SRMRA by about 55.6% on the revenue averagely under various scenarios.

References:

1. Miettinen A P, Nurminen J K.: Energy e_ciency of mobile clients in cloud computing. In: Usenix Conference on Hot Topics in Cloud Computing, pp. 4-4. USENIX Association, Boston, MA (2010)

2. Burgstahler D, Richerzhagen N, Englert F, et al.: Switching Push and Pull: An Energy Efficient Notification Approach. In: IEEE International Conference on Mobile Services, pp. 68-75. IEEE, Anchorage, AK, USA (2014)

3. Ahn S, Lee J, Park S, et al.: Competitive Partial Computation Offloading for Maximizing Energy Efficiency in Mobile Cloud Computing. IEEE Access PP(99),1-1 (2018)

4. Thanapal P, Durai M A S.: A framework for computational offloading to extend the energy of mobile devices in mobile cloud computing. International Journal of Embedded Systems 9(5), 444 (2017)
5. E. Cuervo, A. Balasubramanian, D. Cho, A. Wolman, S. Saroiu, R. Chandra, and P. Bahl.: Maui: making smartphones last longer with code offload. In: Proceedings of the 8th International Conference on Mobile Systems, Applications, and Services, pp. 49-62. ACM MobiSys, San Francisco, CA, USA (2010)
6. Chun B G, Ihm S, Maniatis P, et al.: CloneCloud: elastic execution between mobile device and cloud. In: Conference on Computer Systems, pp. 301-314. ACM, Salzburg, Austria (2011)
7. Khan A U R, Othman M, Madani S A, et al.: A Survey of Mobile Cloud Computing Application Models. IEEE Communications Surveys & Tutorials 16(1), 393-413 (2014)
8. Wu J, Yuen C, Cheung N M, et al.: Enabling Adaptive High-Frame-Rate Video Streaming in Mobile Cloud Gaming Applications. IEEE Transactions on Circuits & Systems for Video Technology 25(12), 1988-2001 (2015)
9. Chunlin L I, Layuan L I.: An Optimization Approach for Utilizing Cloud Services for Mobile Devices in Cloud Environment. Informatica 26(1), 89-110 (2015)
10. Meng S, Wang Y, Miao Z, et al.: Joint optimization of wireless bandwidth and computing resource in cloudlet-based mobile cloud computing environment. Peer-to-Peer Networking and Applications 11(3), 462-472 (2017)
11. Tocz, Klervie, S. Nadjmtehrani.: A Taxonomy for Management and Optimization of Multiple Resources in Edge Computing. CoRR (2018)
12. Chen L, Wu J, Zhang X X, et al.: TARCO: Two-Stage Auction for D2D Relay Aided Computation Resource Allocation in HetNet. IEEE Transactions on Services Computing, doi: 10.1109/TSC.2018.2792024, (2018)
13. Mtibaa A, Fahim A, Harras K A, et al.: Towards resource sharing in mobile device clouds: power balancing across mobile devices. In: ACM SIGCOMM Workshop on Mobile Cloud Computing, pp. 51-56. ACM, Hong Kong, China (2013)
14. Mtibaa A, Harras K A, Fahim A.: Towards Computational O_oading in Mobile Device Clouds. In: IEEE, International Conference on Cloud Computing Technology and Science, pp. 331-338. IEEE, Bristol, UK (2014)

15. Fahim A, Mtibaa A, Harras K A.: Making the case for computational o_oading in mobile device clouds. In: International Conference on Mobile Computing & NETWORKING, pp. 203-205. ACM, New York, NY, USA (2013)
16. Habak K, Shi C, Zegura E W, et al.: Elastic Mobile Device Clouds: Leveraging Mobile Devices to Provide Cloud Computing Services at the Edge. Fog for 5G and IoT (2017)
17. Miluzzo E, Chen Y F.: Vision:mClouds - computing on clouds of mobile devices. In: ACM Workshop on Mobile Cloud Computing and Services, pp. 9-14. ACM, Low Wood Bay, Lake District, UK (2012)
18. Song J, Cui Y, Li M, et al.: Energy-tra_c tradeo_ cooperative o_oading for mobile cloud computing. In: Quality of Service, pp. 284-289. IEEE, Hong Kong, China (2014)
19. Wang X, Chen X, Wu W, et al.: Cooperative Application Execution in Mobile Cloud Computing: A Stackelberg Game Approach. IEEE Communications Letters 20(5), 946-949 (2016)
20. Zaman S, Grosu D.: Combinatorial Auction-Based Allocation of Virtual Machine Instances in Clouds. In: IEEE Second International Conference on Cloud Computing Technology and Science. IEEE Computer Society, pp. 127-134. IEEE, Indianapolis, IN, USA (2010)
21. Shi W, Zhang L, Wu C, et al.: An online auction framework for dynamic resource provisioning in cloud computing. IEEE/ACM Transactions on Networking 24(4), 2060-2073 (2016)
22. Zhu Y, Li B, Li Z.: Truthful spectrum auction design for secondary networks. In: IEEE INFOCOM, pp. 873-881. IEEE, Orlando, FL, USA (2012)
23. Wang X, Huang L, Xu H, et al.: Social Welfare Maximization Auction for Secondary Spectrum Markets: A Long-Term Perspective. In: IEEE International Conference on Sensing, Communication, and NETWORKING, pp. 1-9. IEEE, London, UK (2016)
24. Vickrey, William.: Counterspeculation, Auctions, and Competitive Sealed Tenders. Journal of Finance 16(1), 8-37 (1961)
25. Li H,Wu C, Li Z.: Socially-optimal online spectrum auctions for secondary wireless communication. In: Computer Communications, pp. 2047-2055. IEEE, Kowloon, Hong Kong (2015)

26. Jin, Along, W. Song, and W. Zhuang.: Auction-Based Resource Allocation for Sharing Cloudlets in Mobile Cloud Computing. IEEE Transactions on Emerging Topics in Computing 6(1), 45-57 (2018)
27. Wang X, Chen X, Wu W.: Towards truthful auction mechanisms for task assignment in mobile device clouds. In: IEEE INFOCOM 2017 - IEEE Conference on Computer Communications, pp. 1-9. IEEE, Atlanta, GA, USA (2017)
28. Wang H, Guo S, Cao J, et al.: MELODY: A Long-term Dynamic Quality-Aware Incentive Mechanism for Crowdsourcing. IEEE Transactions on Parallel & Distributed Systems 29(4), 901-914 (2018)

For full version, please refer to http://dx.doi.org/10.1007/978-3-030-05057-3_28 when it is online.